\documentclass[final, twocolumn]{revtex4}
\pdfoutput=1

\usepackage{amssymb}
\usepackage{amsmath}
\usepackage{epsfig}

\newcommand{\kB}{k_{\mathrm{B}}}
\newcommand{\kT}{\kB T}
\newcommand{\Angstrom}{\mathrm{\AA}}
\newcommand{\vek}[1]{\boldsymbol{#1}}          
\newcommand{\dif}{\mathrm{d}}                  
\newcommand{\mean}[1]{\left<#1\right>}
\newcommand{\gmax}{g_{\mathrm{max}}}
\newcommand{\rmin}{r_{\mathrm{min}}}
\newcommand{\Phimin}{\Phi_{\mathrm{min}}}
\newcommand{\gprox}{g_{\mathrm{prox}}}
\newcommand{\rprox}{r_{\mathrm{prox}}}
\def\Real{\hbox{I\kern-.1667em\hbox{R}}}

\newcommand{\abs}[1]{\vert {#1} \vert}

\usepackage[letterpaper,hmargin=0.9in,vmargin=1.1in]{geometry}
            
\begin{document}

\title{Segue Between Favorable and Unfavorable Solvation}
\author{Lutz Maibaum}
\author{David Chandler}
\affiliation{Department of Chemistry, University of California, Berkeley, California 94720}
\date{\today}

\begin{abstract}
Solvation of small and large clusters are studied by simulation, 
considering a range of solvent-solute attractive energy strengths. 
Over a wide range of conditions, both for solvation in the 
Lennard-Jones liquid and in the SPC model of water, it is shown 
that the mean solvent density varies linearly with changes in 
solvent-solute adhesion or attractive energy strength.
This behavior is understood from the perspective 
of Weeks' theory of solvation [{\em Ann. Rev. Phys. Chem.} {\bf 2002}, {\em 53}, 533] and 
supports theories based upon that perspective.
\end{abstract}

\maketitle

\affiliation{Department of Chemistry, University of California, Berkeley, CA
94720-1460}

\section{\label{sec:theory}INTRODUCTION}

\label{intro}Some solute species dissolve in a liquid solvent over a large
range of conditions. Others demix or phase separate from the liquid solvent
even at relatively low solute concentrations. This paper focuses on
mesoscopic and microscopic manifestations of these macroscopic phenomena.
Our motivation is to understand the range of solvation behaviors that are
possible in aqueous solutions when hydrophobic species are altered to
hydrophilic and vice versa. While the motivation concerns this specific
class of solutions, the physical principles involved are relatively general.

The basic theory has been formulated by Weeks~\cite{Weeks02}. One first
considers the effect of the solute in excluding a region of space from the
solute. The average effect is described by the response of the solvent to a
so-called unbalancing potential. Unbalancing arises because the excluded
volume presents a region where solvent molecules lose adhesive forces. In a
homogeneous fluid, these adhesive forces tend to be in balance, often allowing
their effects to be described as a uniform field or chemical potential shift~%
\cite{Widom67, Weeks71}. Introducing a large enough excluded volume, however,
destroys this homogeneity, and the response can be very large, particularly
so if the liquid exists near coexistence with its vapor. Indeed, in the
macroscopic limit, the average liquid density adjacent to excluded regions
bounded by a plane is $\beta p$, where $p$ is the pressure, and $\beta
^{-1}=k_{\mathrm{B}}T$ is temperature times Boltzmann's constant~\cite{Evans88}. In this
limit, therefore, the mean liquid density at the surface is a bulk property.
It exhibits a phase transition reflecting the bulk liquid-vapor phase
transition. Weeks' unbalancing potential captures the effects of this
transition in a mean molecular field approximation.

A typical solute does not exclude liquid from a macroscopic volume.
Nevertheless, the unbalancing potential can be substantial and the response
to it significant, though not singular, provided the solvent-exposed surface
area of the solute has low enough curvature and large enough size. Weeks'
expression for the unbalancing potential shows that its strength is roughly
proportional to the surface area, $A$. Thus, to linear order
in the response,
\begin{equation}
\left\langle \rho \! \left( \mathbf{r}\right) \right\rangle _{0}\thickapprox
a \! \left( \mathbf{r}\right) -A \, b \! \left(\mathbf{r}\right) .  \label{rho0}
\end{equation}
Here, $\rho \! \left( \mathbf{r}\right) $ denotes the solvent density field at
position $\mathbf{r}$, which we take to be outside the excluded volume, and
the pointed brackets indicate equilibrium ensemble average, with the
subscript 0 implying that only excluded volume forces act between solute and
solvent. The function $a \! \left( \mathbf{r}\right) $ and $b\! \left( \mathbf{r}
\right) $ are properties of the solvent and the solute position and shape; $%
a\! \left( \mathbf{r}\right) $ gives the response in the absence of unbalancing
forces; $A\,b\! \left( \mathbf{r}\right) $ is significant in size only when the
liquid solvent is close to phase coexistence with its vapor and $A$ is
comparable to or larger than the surface area of the critical nucleus for
the phase transition; for positions $\mathbf{r}$ that are beyond a bulk
correlation length of the solvent, $a\! \left( \mathbf{r}\right) $ approaches
the bulk liquid density, and $b\! \left( \mathbf{r}\right) $ approaches zero.

Fluctuations around the average and the effects of forces beyond those of
excluded volume are captured in the second step of Weeks' treatment. In this
step, one notes that solvent density fluctuations in a homogeneous liquid
obey Gaussian statistics~\cite{Hummer96, Crooks97, Chandler93}, and assumes this
statistics is more generally valid for all small length scale fluctuations
in a liquid, even in the presence of inhomogeneities. Therefore, if the
solute attracts the solvent molecules with a potential strength
proportional to $\lambda $, the effect of the attraction on the mean solvent
density is linear in $\lambda $, i.e., $\left\langle \rho \! \left( \mathbf{r}
\right) \right\rangle _{\lambda }\thickapprox \left\langle \rho \! \left(
\mathbf{r}\right) \right\rangle _{0}+\lambda \, S\! \left( \mathbf{r}\right) ,$
where $S\! \left( \mathbf{r}\right) $ is the net or integrated response of the
solvent density field at $\mathbf{r}$ to the attractions from the solute. If
these attractions emanate from interaction sites in the solute and are
finite ranged, then $S\! \left( \mathbf{r}\right) $ will be approximately
linear in $A$, leading to
\begin{equation}
\left\langle \rho \! \left( \mathbf{r}\right) \right\rangle _{\lambda
}\thickapprox \left\langle \rho \! \left( \mathbf{r}\right) \right\rangle
_{0}+\lambda \,A\,s\! \left( \mathbf{r}\right) ,  \label{eq:rhoLambda}
\end{equation}
where $s\! \left( \mathbf{r}\right) $ is a property of the solvent and the
functional form of the attractive interaction. Finally, as both $b\! \left(
\mathbf{r}\right) $ and $s\! \left( \mathbf{r}\right) $ describe responses of
the solvent near the solute and per unit exposed area of the solute, one may
expect that integrated or coarse-grained effects of the two are
approximately proportional. That is to say, $\overline{b}\! \left( \mathbf{r}
\right) \thickapprox \lambda ^{*} \, \overline{s}\! \left( \mathbf{r}\right) $,
where the over-bar indicates an integral or coarse-graining over some
specified microscopic length scale. The value of the proportionality
constant, $\lambda ^{*}$, depends upon the specific solvent, the specific
solute-solvent interactions, and the specific coarse-graining perscription.
Adopting this proportionality gives
\begin{equation}
\left\langle \overline{\rho }\! \left( \mathbf{r}\right) \right\rangle
_{\lambda }\thickapprox \overline{a}\! \left( \mathbf{r}\right) +\left( \lambda
-\lambda ^{*}\right) \,A\,\overline{s}\! \left( \mathbf{r}\right) .
\label{rhoFinal}
\end{equation}

Equations (\ref{rho0}), (\ref{eq:rhoLambda}) and (\ref{rhoFinal}) are not
generally true, as they do not properly account for saturation in either the
limit of very large $A$ or the limit of very large $\lambda .$ The point of
this paper, however, is that these equations are good approximations in a
variety of physically pertinent circumstances. We demonstrate their accuracy
in the next section by presenting results of computer simulation
calculations of solvent density fields and solvation free energies for
solute clusters of various sizes and solvent-solute attractive energy
strengths, considering both the Lennard-Jones liquid solvent and water
solvent. The results indicate that variations from favorable to unfavorable
solvation, i.e., the changes in solvation in passing from large $\lambda$
to small $\lambda$, occur smoothly and in ways that are easy to anticipate.
Further, the results provide confidence in theories built from Weeks'
perspective. These implications are discussed in Section III.

\section{Simulation Results}

To elucidate the response of a liquid to the presence of a solute of general
shape and interaction strength, we consider a series of models for both
solutes and solvent of increasing complexity.

\subsection{Spherical solutes in Lennard--Jones solvent}

First we consider the case of a spherical solute immersed in the
Lennard--Jones solvent, i.e., a fluid of particles that interact through the
pair potential
\begin{equation}
w_{\mathrm{LJ}} (r) = 4 \varepsilon \left[ \left(\frac{\sigma}{r}\right)^{12}
  - \left(\frac{\sigma}{r}\right)^6 \right] .
\end{equation}
In this equation $r$ denotes the distance between two solvent particles, and
$\varepsilon$ and $\sigma$ define the energy and length scales of the simulation.
The solvent--solvent potentials are shifted so as to be continuous and
zero beyond the cut-off distance $r_{\mathrm{c}}=2.5\sigma $. The reduced
thermodynamic state conditions for the solvent are 
 $k_{\mathrm{B}}T/\varepsilon = 0.85$, $p \sigma^3 / \varepsilon = 0.022$ and
 $\rho \sigma ^{3}= 0.7,$ where $\rho $
 is the bulk density. These conditions place this Lennard-Jones liquid
 solvent at or very close to liquid-vapor coexistence~\cite{Smit92}. 

The solute is a sphere of radius $R$, and its interaction with the solvent is
that of a sphere  consisting of a Lennard--Jones material with homogenous
density equal to that of the solvent. A solvent particle at distance $r$
from the solute center has the potential energy\cite{Huang02}
\begin{multline}
\Phi (r) =
 \pi \varepsilon \rho \sigma^3 \left[
\frac{4}{5}\sigma^9 \left(
        \frac{1}{8 r r_+^8}
        - \frac{1}{9 r_+^9}
        - \frac{1}{8 r r_-^8}
        + \frac{1}{9 r_-^9}
\right) \right. \\
- 
\left. 2 \sigma^3 \left(
        \frac{1}{2 r r_+^2}
        - \frac{1}{3 r_+^3}
        - \frac{1}{2 r r_-^2}
        + \frac{1}{3 r_-^3}
\right)
\right] ,
\label{eq:HomPotential}
\end{multline}
where $r_\pm  = r \pm R$, if $r \ge R$, and $\phi(r) = \infty$ otherwise. This
potential is minimal at a distance $\rmin$, and its minimum value
$\Phimin = \Phi(\rmin)$ depends on the solute size $R$.

To systematically investigate the role of attractive interactions, we
employ the scheme introduced by Weeks, Chandler and Andersen (WCA)
to partition the energy \eqref{eq:HomPotential} into the short--ranged repulsive
and the long--ranged attractive part~\cite{Weeks71},
\begin{equation}
  \Phi_\lambda (r) = \Phi_0 (r) + \lambda \Phi_1 (r) ,
\label{eq:HomPotentialPartitioned}
\end{equation}
where
\begin{equation}
  \Phi_0(r) = 
\left\{
\begin{array}{cl}
 \Phi(r) - \Phi(r_{\mathrm{min}}) & \mathrm{if} \; r \le r_{\mathrm{min}} \\ 
 0      & \mathrm{if} \; r > r_{\mathrm{min}}
\end{array} 
\right. 
\end{equation}
and
\begin{equation}
  \Phi_1(r) = 
\left\{
\begin{array}{cl}
 \Phi(r_{\mathrm{min}})  & \mathrm{if} \; r \le r_{\mathrm{min}} \\ 
 \Phi(r)    & \mathrm{if} \; r > r_{\mathrm{min}} .
\end{array} 
\right. 
\end{equation}
The parameter $\lambda$ controls the strength of the attractive
solute--solvent interactions.

We performed computer simulations using the Monte Carlo method at fixed
pressure and temperature for different values of the solute size $R$ and
interaction strength $\lambda$. The number $N$ of solvent particles ranged from
864 to 8192 depending the size of the solute. The solute--solvent interaction
\eqref{eq:HomPotentialPartitioned} was truncated and shifted at a distance $r
= R + 2.5 \sigma$. We calculate the average reduced solvent density at position $\vek{r}$,
\begin{equation}
  g (\vek{r}) = \mean{ \rho (\vek{r})} / \rho ,
\end{equation}
in an ensemble where the solute position is fixed at $\vek{r} = 0$. Due to the
isotropic nature of the solute this density depends only on the distance $r =
\abs{\vek{r}}$ from the solute center.

\begin{figure}[t]
\resizebox{\columnwidth}{!}{
\includegraphics{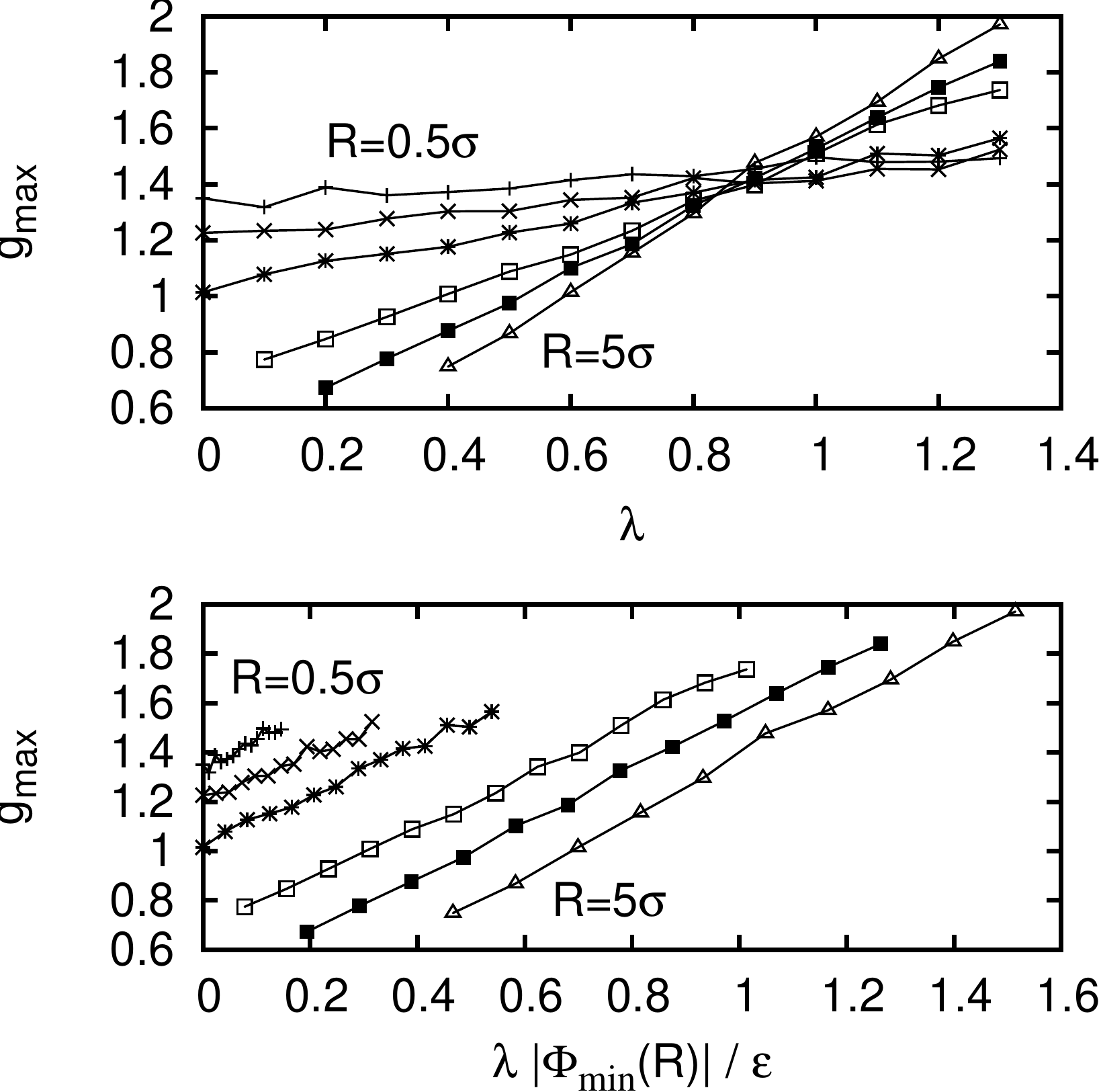}
}
\caption{\label{fig:homsphere}
Magnitude of the first peak of the solvent density distribution function
around homogeneous Lennard--Jones solutes of radii $R/\sigma = 0.5, 0.7, 1, 2,
3, 5$, plotted as a function of the bare attraction strength $\lambda$ (top)
and the minimum of the solvent--solute potential (bottom). For large solute sizes
and small values of $\lambda$ the reduced density distribution function did not have a
well--defined first peak, and no data is shown for those systems.
}
\end{figure}

We find that the reduced density $g(r)$ has generally a well--defined first
peak corresponding to the first solvation shell of the solute. The height of this
peak, denoted $\gmax$, is shown in the top panel of Fig. \ref{fig:homsphere} for different
values of $R$ and $\lambda$. This graph encompasses many of the important
effects of our understanding of solvophobic and solvophilic solvation.

First, for a fixed value of $\lambda$ less than $0.9$, the solvent density
near the solute decreases with increasing solute radius. It has been observed
previously that the contact density near a hard sphere solute decreases from
values above the bulk liquid density at small solute sizes to the equilibrium
vapor density at large solute sizes, both for the Lennard--Jones solvent
\cite{Huang00} and water \cite{Huang02}. The creation of a liquid--vapor
interface near large enough solutes is the basis of our understanding of the
hydrophobic effect \cite{Chandler05}. Our simulations show that this behavior
persists for solutes with weak but finite attraction strengths $\lambda$.

Second, for any given solute size $R$, the density near the solute increases
approximately linearly with the attraction strength $\lambda$. This smooth
response of the density to an attractive field over a wide range of
interactions strengths shows that the solvent properties do not change
significantly as $\lambda$ increases. In particular it is consistent with
Gaussian density fluctuations for all attraction strengths.

Third, it is apparent that the slope $\partial \gmax / \partial \lambda$,
while independend of $\lambda$, depends on the radius $R$ of the solute. In
particular, there is a value $\lambda^* \approx 0.9$ of the interaction
strength where the mean density itself seems to be independent of the solute
size. A similar result is suggested by Ashbaugh and Paulaitis for the density
near methane clusters in water~\cite{Ashbaugh01}. This behavior can be easily
understood within the framework introduced in Section \ref{sec:theory}. The
numerical value of $\lambda$ by itself is not an accurate parameter to describe the strength
of the solute--solvent interactions. For the special case of the homogeneous
Lennard--Jones solute, a much more reliable indicator is the magnitude of the
minimum in the potential energy, $\lambda \abs{\Phimin (R)}$. The bottom panel
of Fig. \ref{fig:homsphere} shows that the density response is the same for
all solute sizes if the control parameter is chosen to be the actual magnitude
of the solute--solvent attractions.

Our results for this simple model solute serve as a useful illustration for the
theory outlined in Section \ref{sec:theory}. We now turn to more complicated
solutes, where the lack of isotropy demands the analysis of more complicated
density distribution functions. While these complications can obscur the
underlying physics, we will see that the basic picture of solvation presented
here remains correct.

\subsection{Cluster solutes in Lennard--Jones solvent}

\begin{figure}[t]
\resizebox{\columnwidth}{!}{
\includegraphics{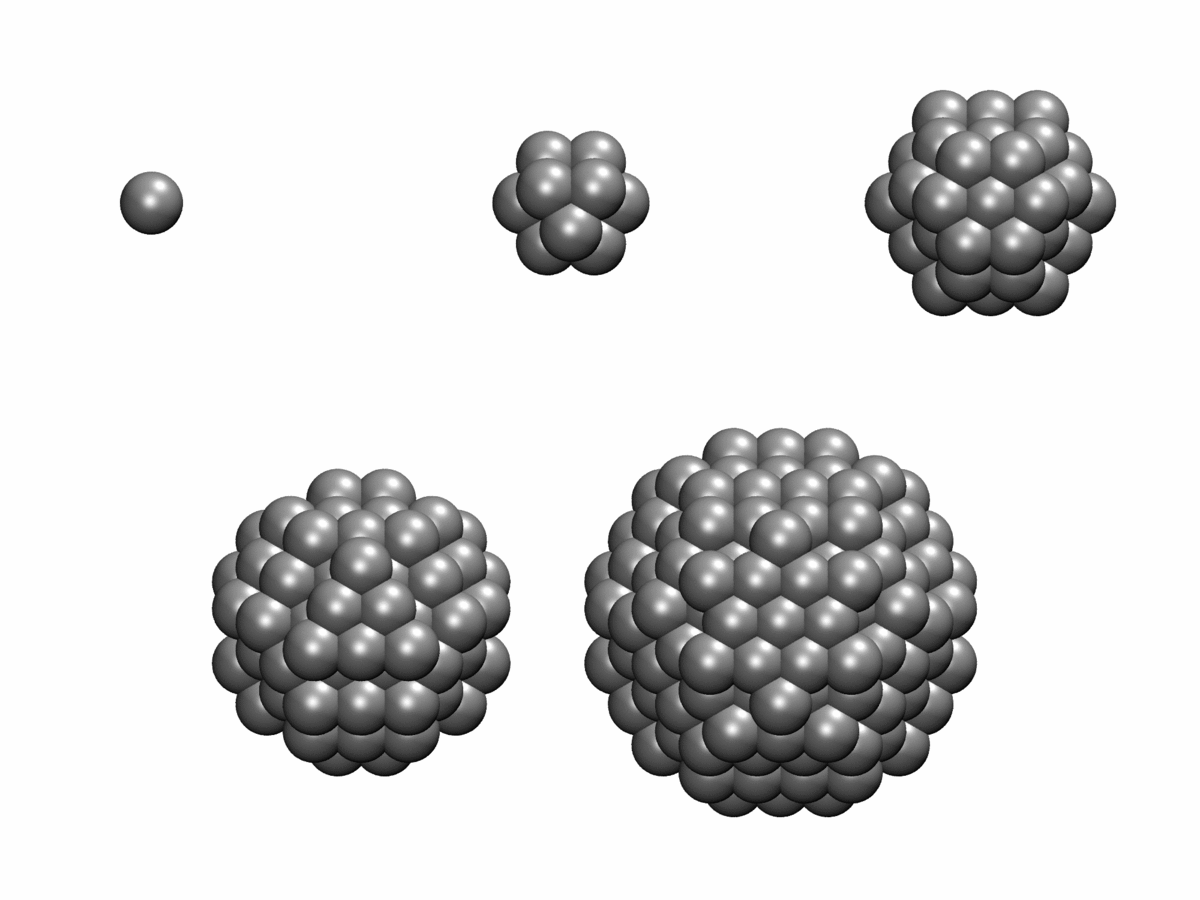}
}
\caption{\label{fig:clusters}
Space filling view of the solute clusters with particle numbers $n=1,13,57,135,305$. The
clusters are spherical cuts of a hexagonally close--packed crystal.
}
\end{figure}

We consider clusters of $n$ particles solvated in the same Lennard--Jones
fluid as before. The clusters are the model solutes of Ashbaugh and
Paulaitis~\cite{Ashbaugh01}. Space filling views are shown in Figure \ref{fig:clusters}. The clusters are
composed of $n=1,13,57,135,$ and $305$ hexagonally close packed particles
with nearest neighbor distance $d = 2^{1/6} \sigma$. The solute interacts with the solvent
molecules with the potential energy
\begin{equation}
W_\lambda = \sum_{i=1}^{N} \sum_{\alpha=1}^{n} w_\lambda \left(
  \abs{\vek{r}_\alpha - \vek{R}_i} \right)
\end{equation}
where $\vek{r}_\alpha$ is the position of the $\alpha$th cluster particle,
$\vek{R}_i$ is the position of the 
$i$th solvent particle, and
\begin{equation}
  w_\lambda (r) = w_0 (r) + \lambda w_1 (r) .
\end{equation}
Here, $w_0 (r)$ and $w_1 (r)$ are, respectively, the repulsive branch and
the attractive branch of the Lennard--Jones potential~\cite{Weeks71},
\begin{equation}
  w_0 (r) = 
\left\{
\begin{array}{cl}
4 \varepsilon_{\mathrm{s}} \left[ \left(\sigma_{\mathrm{s}} / r\right)^{12}
  - \left(\sigma_{\mathrm{s}} / r\right)^6  + \frac{1}{4} \right]
 & \mathrm{if} \; r \le 2^{1/6} \sigma_{\mathrm{s}} \\ 
 0 & \mathrm{if} \; r > 2^{1/6} \sigma_{\mathrm{s}}
\end{array} 
\right. 
\end{equation}
and
\begin{equation}
  w_1 (r) = 
\left\{
\begin{array}{cl}
 -\varepsilon_{\mathrm{s}} & \mathrm{if} \; r \le 2^{1/6} \sigma_{\mathrm{s}} \\ 
 4 \varepsilon_{\mathrm{s}} \left[ \left(\sigma_{\mathrm{s}} / r\right)^{12}
  - \left(\sigma_{\mathrm{s}} / r\right)^6   \right] & \mathrm{if} \; r > 2^{1/6} \sigma_{\mathrm{s}}
\end{array} 
\right. .
\end{equation}
We choose $\sigma_{\mathrm{s}} = \sigma$ and $\varepsilon_{\mathrm{s}} = \varepsilon$ so that the
interaction between a cluster particle and a solvent particle is the same as
that between two solvent particles at $\lambda = 1$.

The $n=1$ solute is spherically symmetric, and the mean solvent density
near it is described by the isotropic radial distribution function $g(r)$ as
before. For the clusters with $n>1$, however, specification of the 
mean solvent density requires many more variables than $r$, and it is natural
to consider projections of the mean density. One such
projection is the so-called proximal distribution $\gprox(r)$, defined as
the reduced density averaged over the surface~\cite{Ashbaugh01, Mehrotra80}
\begin{equation}
  \left\{ \vek{r}' \in \Real^3 : \rprox \left( \vek{r}' \right) = r  \right\} ,
\label{eq:proximalsurface1}
\end{equation}
where
\begin{equation}
  \rprox \left( \vek{r} \right) = \min \left\{ \abs{\vek{r} - \vek{r}_\alpha} : \alpha = 1, \ldots, n \right\}
\label{eq:proximalsurface2}
\end{equation}
is the distance from a point $\vek{r}$ to the nearest cluster particle. While explicitly a function
of only one length, $r$, it is implicitly a multi--point distribution
function~\cite{footnotegprox},
except in the case $n=1$. For that special case  
$\gprox(r)$ is the standard radial distribution function $g(r)$.

\begin{figure}[t]
\resizebox{\columnwidth}{!}{
\includegraphics{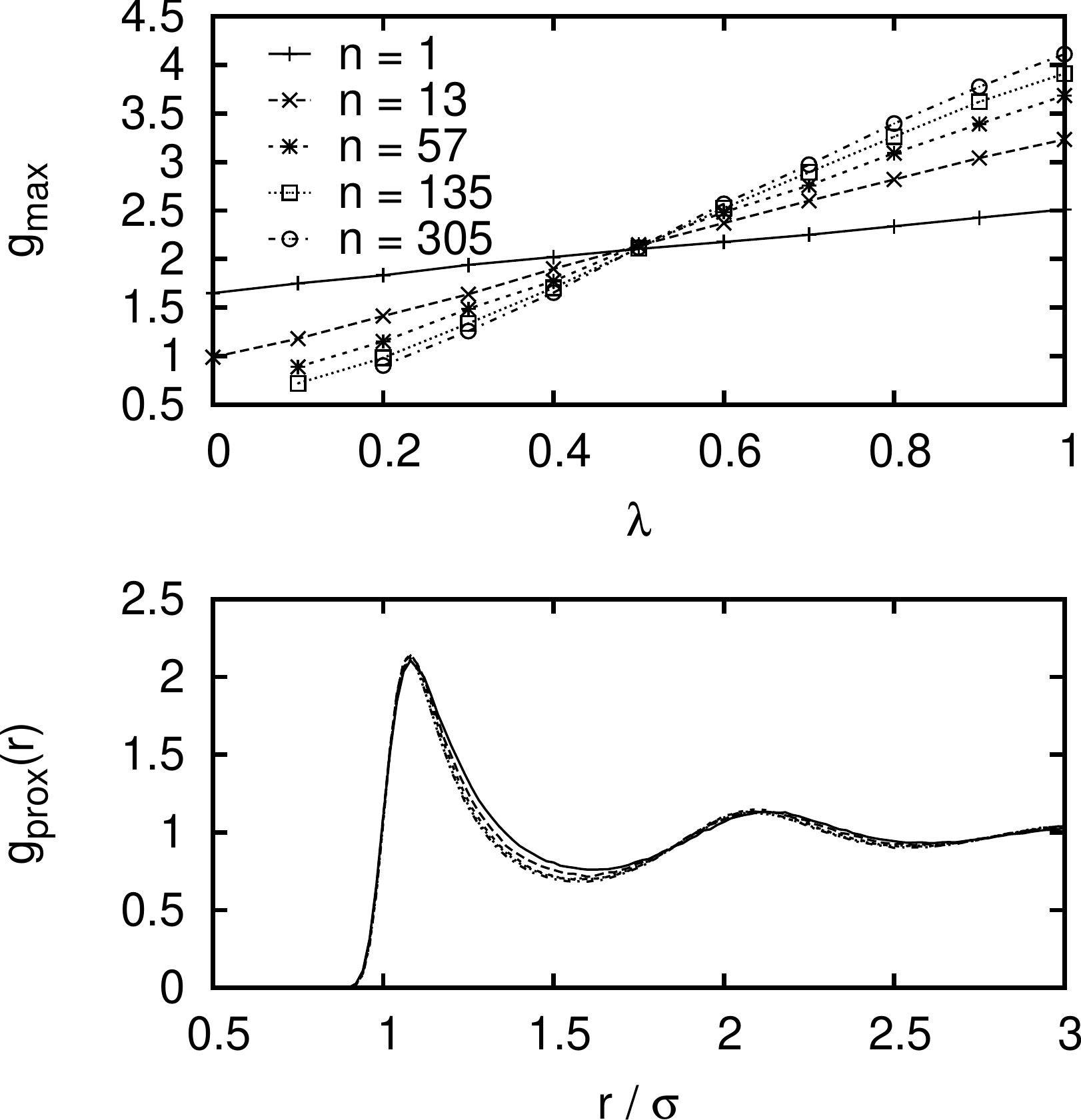}
}
\caption{\label{fig:clprox1}
Proximal solvent density distribution function around clusters of Lennard--Jones particles
immersed in the Lennard--Jones fluid. (top) Height of the first peak in
$\gprox(r)$ for different cluster sizes $n$ and attraction strengths
$\lambda$. (bottom) Proximal distribution functions for $\lambda = \lambda^* = 0.5$.
}
\end{figure}

Figure \ref{fig:clprox1} illustrates the behavior of the proximal distribution function for
the model solutes in the Lennard--Jones solvent. Ashbaugh and Paulaitis
proposed the height $\gmax$ of the first peak in $\gprox (r)$ as a meaningful
measure of the solvent density near arbitrary solutes. In the top panel we
show that the behavior of this observable is qualitatively the same as that
shown in Fig. \ref{fig:homsphere} for isotropic solutes. The response of the mean
solvent density to changes in solute-solvent attraction is proportional to
that change. This result is as expected according to Eq.~\eqref{eq:rhoLambda}.
Another important point is that this proportionality leads to a coincidence
of curves at $\lambda =\lambda ^{*}$, where in this case $\lambda 
^{*}\approx 0.5$. 
The existence of a coincident value of $\lambda $ is anticipated in Eq.~\eqref{rhoFinal}.

\begin{figure}[t]
\resizebox{\columnwidth}{!}{
\includegraphics{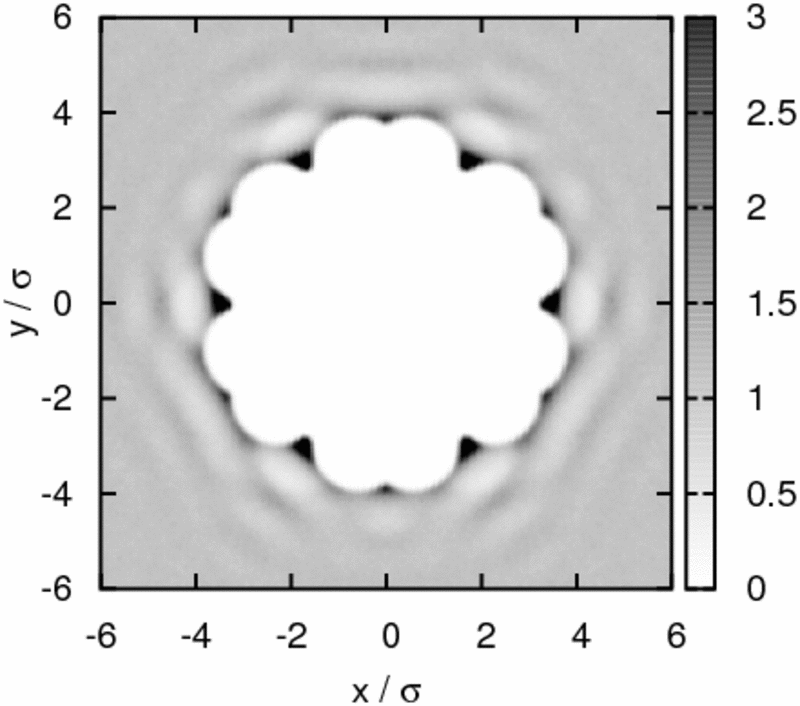}
}
\caption{\label{fig:ljcl135densitymap}
Reduced solvent density around the $n=135$ solute
cluster immersed in the Lennard--Jones fluid
at $\lambda = \lambda^* = 0.5$. Shown is a slice through the solute center of
width $0.1 \sigma$ with resolution $0.05 \sigma$, demonstrating that the solvent density does not
follow the shape of the solute surface.
}
\end{figure}

Ashbaugh and Paulaitis found in their study of methane clusters solvated in
water that at $\lambda^*$ not only the heights of the first peak in $\gprox
(r)$ coincide for different cluster sizes, but that the
proximal distribution functions themselves are very similar. The bottom panel of
Fig. \ref{fig:clprox1} establishes a similar result for our model system. From
this behavior, Ashbaugh and Paulaitis conclude that the solvation environment of a monomeric methane particle
is essentially the same as that of methane confined to a cluster of arbitrary
size. Fig. \ref{fig:ljcl135densitymap} shows that this conclusion is not
justified. The contour lines of
constant density do not coincide with the surfaces of constant proximal
distance for the $n=135$ cluster. In particular, the crevices of the solute, where the attraction
strength is large, contain a much higher solvent density than the proximal
distribution function suggests. No such behavior exists for the monomeric
solute, where the density profile is isotropic.

As the strength of the attactive solute--solvent interaction increases, the
clusters lose their hydrophobic character until eventually they become
hydrophilic. To quantify this process we calculate the solvation free energy,
or excess chemical potential, as a function of $\lambda$, which is given by
\begin{equation}
  \Delta \mu_{\lambda } = \Delta \mu _{0}+\int_{0}^{\lambda } \dif \lambda' 
\mean{W_1 - W_0}_{\lambda'}
\label{eq:DeltaMu}
\end{equation}
where $\mean{\dots}_{\lambda}$ indicates ensemble average
with solute--solvent attractive force strength $\lambda$.

Solvation free energies due to excluded volume are approximately
proportional to the excluded volume for small solutes. Further, at
large excluded volumes with small surface curvature and no attractive
solvent-solute forces, solvent density is depleted near the solute.
Therefore, for the solutes we have considered, $\Delta \mu _{0}$ can be well
approximated by the solvation free energy for a hard sphere with the same
excluded volume as the solute clusters~\cite{Huang00}. Here we define the
excluded volume as those spatial regions where the solute--induced field is
greater than the thermal energy $\kT$.

\begin{figure}[t]
\resizebox{\columnwidth}{!}{
\includegraphics{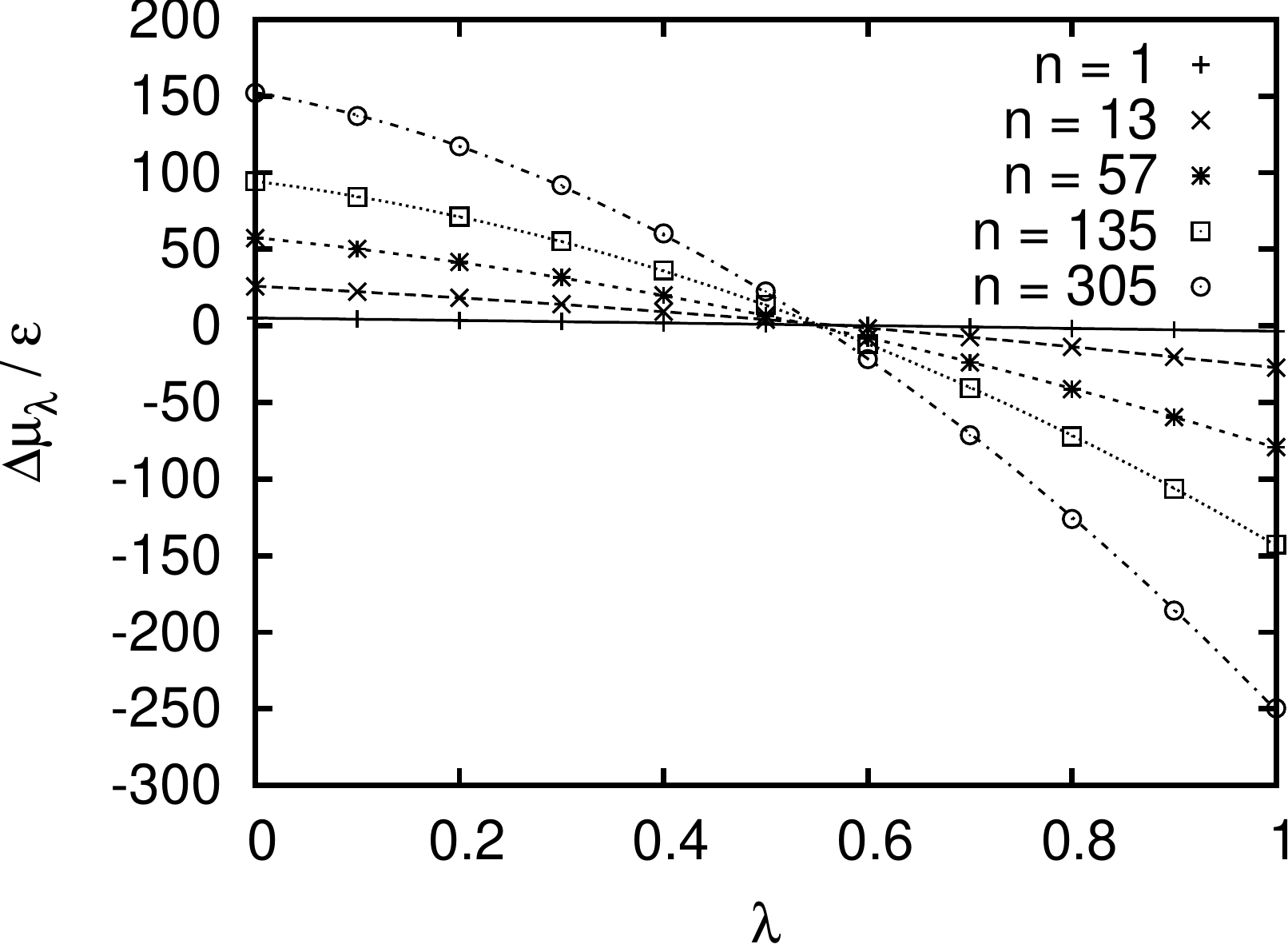}
}
\caption{\label{fig:ljclsolvationfreeenergy}
Excess solvation energy of solute clusters in the Lennard--Jones fluid.
The symbols are calculated from simulation using Eq. \eqref{eq:DeltaMu}, and the
lines are quadratic fits through the data.
}
\end{figure}

The result of this calculation is shown in
Fig. \ref{fig:ljclsolvationfreeenergy}. Over the range of interaction strengths
considered, the solvation free energies change from strongly positive to
negative. In other words, at large values of $\lambda$ the clusters are not
solvophobic. This transition occurs smoothly and regularly. The shape of these
curves is 
understood, because the attractive energy is approximately proportional to
$\lambda$ times an integral over $\mean{\rho (\vek{r})}_\lambda$, and the
latter is a linear function of $\lambda$, as noted in Eq.~\eqref{eq:rhoLambda}.

\subsection{Cluster solutes in SPC water}

In another series of calculations, the solvent is the SPC model
for liquid water \cite{Berendsen81}, and the solutes are clusters of methane
molecules in the united atom approximation of
Ref. \onlinecite{Jorgensen84}. As in Ref. \onlinecite{Ashbaugh01} these
clusters are hexagonally close--packed with a particle spacing of $d = 4.19
\Angstrom$. The methane molecules interact with the water oxygens through a
Lennard--Jones potential with $\sigma = 3.448 \Angstrom$ and $\varepsilon =
0.895 \mathrm{kJ / mol}$, which was truncated and shifted at a distance of
$9 \Angstrom$. As before we use the WCA partition of the interaction potential
to obtain control over the strength of the attractions, given by the parameter
$\lambda$. Electrostatic interactions were calculated using Ewald
summation. Molecular dynamics simulations were performed using the LAMMPS
program \cite{Plimpton95, LAMMPS} at temperature $T = 298 \mathrm{K}$ and a box size
chosen to give the correct bulk density \cite{Ashbaugh01}. 

\begin{figure}[t]
\resizebox{\columnwidth}{!}{
\includegraphics{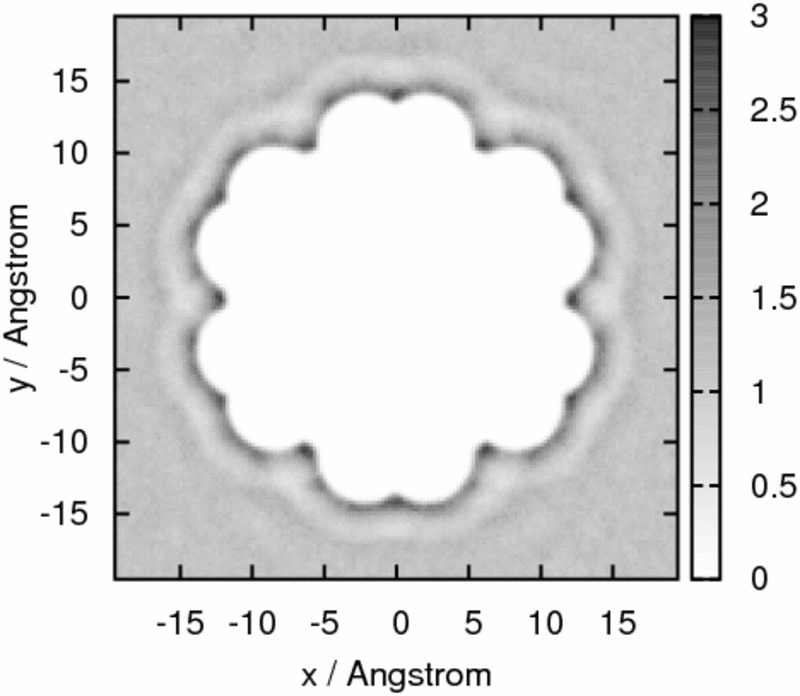}
}
\caption{\label{fig:cl135l1densitymap}
Reduced water density around the $n=135$ solute cluster at $\lambda =
\lambda^* = 1$. Shown is a slice through the solute center of width $0.5
\Angstrom$ with resolution $0.25 \Angstrom$.
To a good approximation the solvent density follows the shape of the solute cluster.
}
\end{figure}

As noted above, Ashbaugh and Paulaitis~\cite{Ashbaugh01} found that the proximal solvent
density distribution functions are very similar for different cluster sizes 
at the specific attraction strength $\lambda^* \approx 1$, from which they concluded that the
solvent density near a cluster of solute particles is basically the same as
that near a single solute particle. In Fig. \ref{fig:cl135l1densitymap} we
show the reduced water density near the $n = 135$ solute with an interaction
strength $\lambda = \lambda^*$. For this specific system the solvent density
is indeed approximately constant along contours of constant proximal distance,
so that
\begin{equation}
\mean{\rho (\vek{r})} \approx \rho \, \gprox (\rprox (\vek{r}))
\end{equation}
is a rather good approximation for the range of parameters considered. However, this agreement is not exact, as
deviations are visible in the crevices. In these regions adjacent to the
cluster, the solvent interacts with many solute particles, and the attractive potential well
reaches values as deep as $-5 \mathrm{kJ/mol}$. 
This value is to be compared with the
average energy of a water molecule in the bulk fluid, which is approximately
$40 \mathrm{kJ/mol}$~\cite{Svishchev93}. Thus, the cluster crevices are
approaching hydrophilic character and can compensate for a fraction of the average water--water
interaction energy.

\begin{figure}[t]
\resizebox{\columnwidth}{!}{
\includegraphics{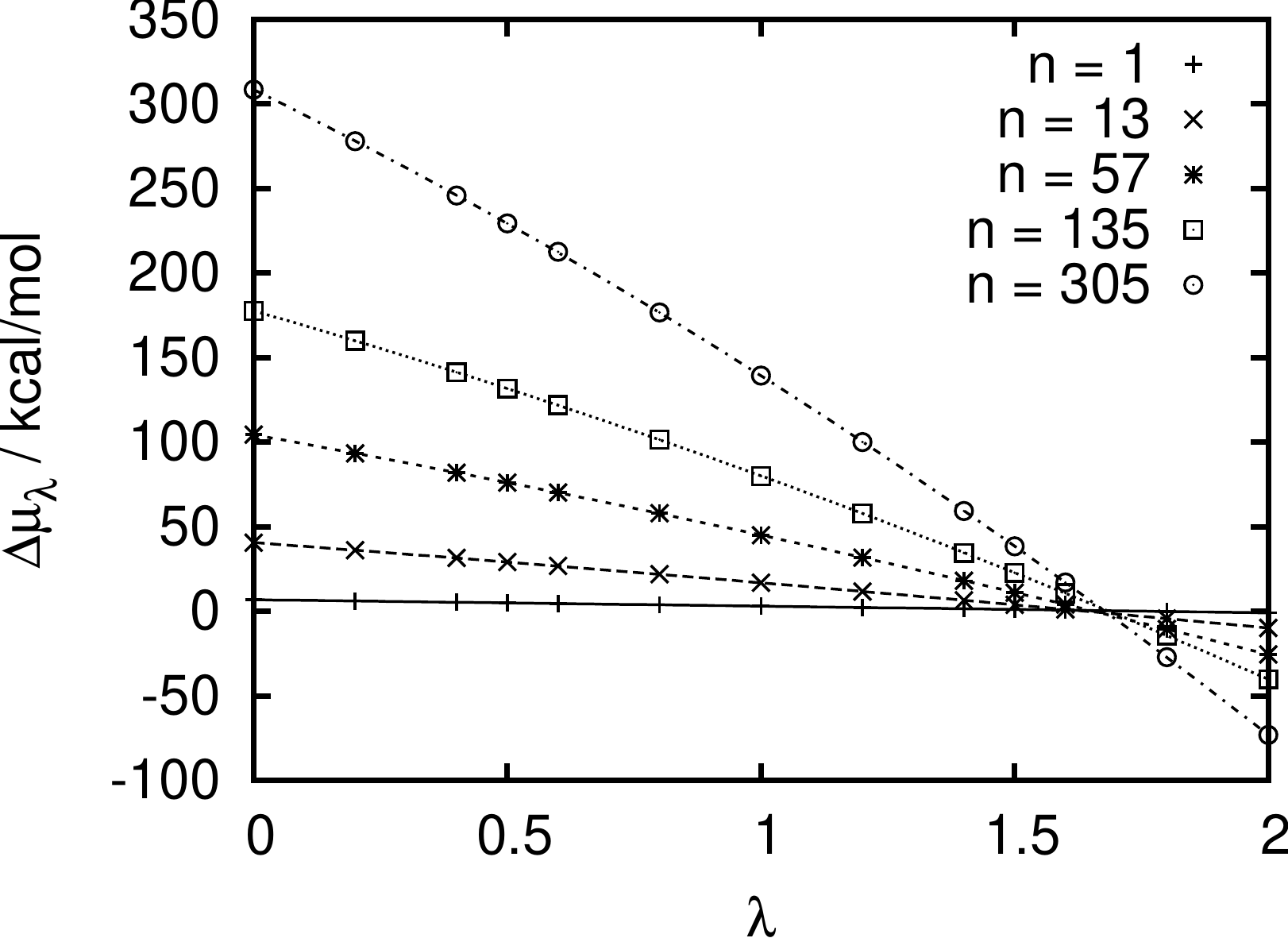}
}
\caption{\label{fig:methanewatersolvationfreeenergy}
Excess solvation energy of methane clusters in water. The symbols are computed
from simulation with the aid of Eq.~\eqref{eq:DeltaMu} and known        
solvation free energies of hard sphere solutes in water~\cite{Huang01}. The lines are
quadratic fits to the data.
}
\end{figure}

Fig. \ref{fig:methanewatersolvationfreeenergy} shows the solvation free energy
of the methane clusters in water, calculated using \eqref{eq:DeltaMu} as
before. Again the clusters show a smooth transition from hydrophobic solvation
at small values of $\lambda$ to hydrophilic solvation at large attraction
strengths. For this particular system, the crossover between these two regimes occurs at values of
$\lambda$ greater than $\lambda^*$.

\section{Discussion}

The results of the previous section show that the role of solvent-solute
attractions on solvation is significant but simple. In particular, while
fluctuations of solvent density are affected by the shape and size of the
solute and the thermodynamic state of the solvent, the mean solvent density
changes only linearly with changes in the strength of adhesion 
between solute and solvent. If, for example, a
solute is small, the liquid solvent will be constrained to be outside the
solute's excluded volume but otherwise behave as if it were bulk. This
conception is precisely the physics that underlies the
Pratt-Chandler theory of hydrophobicity~\cite{Pratt77,Chandler93} and its
contemporary variant~\cite{Hummer96}. On the other hand, if the solute's girth
is such to exclude a volume with large low--curvature surface area,
fluctuations of the solvent are changed to reflect the formation of an
interface~\cite{Weeks02, Lum99, Chandler05}. To the extent there is no
solvent-solute adhesion, the interface is essentially the same as that
between liquid and vapor~\cite{Stillinger73}. The role of attractions between
the solute and solvent is not to change the fluctuations, but to affect the
average solvent density and the position of this interface. For strong enough
attractions, the effect is to pin the interface at specific positions, as
seen in Fig. \ref{fig:ljcl135densitymap}, and to a lesser extent in Fig. \ref{fig:cl135l1densitymap}.

This picture of the role of attractions in the context of hydrophobic
solvation has been discussed before~\cite{Huang02, Chandler05}. 
Simulation results of Werder et al. on the hydration of several 
different models of
graphite in water~\cite{Werder03} can be viewed as illustrations of 
that role. Namely, Werder et al. establish linear
trends for hydration energies as functions of
solvent-solute attractive energy strength. The relative range of parameters is
smaller in that work than in ours, probably explaining why curvature in the
linear trends is not apparent.
Along with demonstrating
regularity and simplicity, Werder et al. also show that the trends are
significant in that modest potential parameter changes result in models of
``graphite'' that change from hydrophobic to hydrophilic. In one model,
Werder et al. consider a Lennard-Jones potential between graphite carbon
atoms and water oxygen atoms with energy and length parameters $\varepsilon
_{\mathrm{CO}}=0.39$~kJ/mol and $\sigma _{\mathrm{CO}}=0.32~$nm. They 
establish that in this model, line tension is positive and the 
contact angle
between water and a planar graphite surface is finite. In contrast, a 
similar model of
a presumed hydrophobic surface, but with $\varepsilon 
_{\mathrm{CO}}=0.48$~kJ/mol
and $\sigma _{\mathrm{CO}}=0.33~$nm yields a negative surface tension
between water and the surface~\cite{Choudhury05,footnotePettit}.
This latter surface is therefore not hydrophobic.

An important consequence of the simplicity we emphasize regards the entropy
of solvation. To the extent that density fluctuations obey Gaussian
statistics (i.e., the mean density responds linearly to forces of adhesion),
the entropy of solvation is independent of adhesive energy strength. This
idea is used in Ref.~\onlinecite{Huang00b} to replace the hydration entropy
of a protein with that of a hard sphere occupying the same volume, even
though the surface of the folded protein is largely hydrophilic. By
demonstrating linear trends for the mean density, our work here provides
explicit justification for this replacement.

\begin{acknowledgments}
This research has been supported in its early stages by the National Science
Foundation and subsequently by U.S. Department of Energy grant No. DE-FG03-87ER13793.

\end{acknowledgments}


\begin{thebibliography}{99}

\bibitem{Weeks02}
Weeks, J.~D. {\em Ann. Rev. Phys. Chem.} {\bf 2002}, {\em 53}, 533.

\bibitem{Widom67}
Widom, B. {\em Science} {\bf 1967}, {\em 157}, 375.

\bibitem{Weeks71}
Weeks, J.~D.; Chandler, D.; Andersen, H.~C. {\em J. Chem. Phys.} {\bf 1971},
  {\em 54}, 5237.

\bibitem{Evans88}
Evans, R. In {\em Liquids at interfaces}; Charvolin, J., Joanny, J.~F.,
  Zinn-Justin, J., Eds.;
\newblock North--Holland, 1988.

\bibitem{Hummer96}
Hummer, G.; Garde, S.; Garcia, A.~E.; Pohorille, A.; Pratt, L.~R. {\em Proc.
  Natl. Acad. Sci. USA} {\bf 1996}, {\em 93}, 8951.

\bibitem{Crooks97}
Crooks, G.~E.; Chandler, D. {\em Phys. Rev. E} {\bf 1997}, {\em 56}, 4217.

\bibitem{Chandler93}
Chandler, D. {\em Phys. Rev. E} {\bf 1993}, {\em 48}, 2898.

\bibitem{Smit92}
Smit, B. {\em J. Chem. Phys.} {\bf 1992}, {\em 96}, 8639.

\bibitem{Huang02}
Huang, D.~M.; Chandler, D. {\em J. Phys. Chem. B} {\bf 2002}, {\em 106},
  2047.

\bibitem{Huang00}
Huang, D.~M.; Chandler, D. {\em Phys. Rev. E} {\bf 2000}, {\em 61}, 1501.

\bibitem{Chandler05}
Chandler, D. {\em Nature} {\bf 2005}, {\em 437}, 640.

\bibitem{Ashbaugh01}
Ashbaugh, H.~S.; Paulaitis, M.~E. {\em J. Am. Chem. Soc.} {\bf 2001}, {\em
  123}, 10721.

\bibitem{Mehrotra80}
Mehrotra, P.~K.; Beveridge, D.~L. {\em J. Am. Chem. Soc.} {\bf 1980}, {\em
  102}, 4287.

\bibitem{footnotegprox}
  Equivalent to eqs. \eqref{eq:proximalsurface1} and \eqref{eq:proximalsurface2},
  the proximal distribution function can be written as $
  \Omega_1 \, r^2 \rho \gprox(r) = \mean{
    \sum_{i=1}^N \delta \left( r- \abs{\vek{R}_i} \right)
    \prod_{\alpha=2}^N \theta \left( \abs{\vek{R}_i - \vek{r}_\alpha} - \abs{\vek{R}_i} \right)
  }
  $. The equilibrium average is performed with solute site $1$ at the origin,
  i.e., $\vek{r}_1 = 0$, this site being at the surface of the cluster and with
  subtended solid angle $\Omega_1$, $\delta (x)$ is the Dirac delta function, and 
  $\theta (x)$ is the unit Heaviside function. The quantity $\rho \gprox(r)$ is the
  mean density of solvent that is closest to and within a shell of radius $r$
  from cluster particle $\alpha = 1$.

\bibitem{Berendsen81}
Berendsen, H. J.~C.; Postma, J. P.~M.; van Gunsteren, W.~F.; Hermans, J. In
  Pullman, B., Ed., {\em Intermolecular forces: Proceedings of the fourteenth
  Jerusalem symposium on quantum chemistry and biochemistry}, page 331,
  Dordrecht, 1981. Reidel.

\bibitem{Jorgensen84}
Jorgensen, W.~L.; Madura, J.~D.; Swenson, C.~J. {\em J. Am. Chem. Soc.} {\bf
  1984}, {\em 106}, 6638.

\bibitem{Plimpton95}
Plimpton, S.~J. {\em J. Comp. Phys.} {\bf 1995}, {\em 117}, 1.

\bibitem{LAMMPS}
{LAMMPS} -- {L}arge--scale {A}tomic / {M}olecular {M}assively {P}arallel
  {S}imulator. Plimpton, S.~J. http://lammps.sandia.gov.

\bibitem{Svishchev93}
Svishchev, I.~M.; Kusalik, P.~G. {\em J. Chem. Phys.} {\bf 1993}, {\em 99},
  3049.

\bibitem{Pratt77}
Pratt, L.~R.; Chandler, D. {\em J. Chem. Phys.} {\bf 1977}, {\em 67}, 3683.

\bibitem{Lum99}
Lum, K.; Chandler, D.; Weeks, J.~D. {\em J. Phys. Chem. B} {\bf 1999}, {\em
  103}, 4570.

\bibitem{Stillinger73}
Stillinger, F.~H. {\em J. Sol. Chem.} {\bf 1973}, {\em 2}, 141.

\bibitem{Werder03}
Werder, T.; Walther, J.~H.; Jaffe, R.~L.; Halicioglu, T.; Koumoutsakos, P. {\em
  J. Phys. Chem. B} {\bf 2003}, {\em 107}, 1345.

\bibitem{Choudhury05}
Choudhury, H.; Pettitt, B.~M. {\em J. Am. Chem. Soc.} {\bf 2005}, {\em
  127}, 3556.

\bibitem{footnotePettit}
In Ref. \onlinecite{Werder03}, Werder et al. do not report the oil--water
surface tension for these particular parameters, but Choudhury and Pettitt, in effect,
do so in Ref. \onlinecite{Choudhury05}. Specifically,
for carbon-carbon potential parameters, the authors of \cite{Choudhury05} cite $%
\varepsilon _{\mathrm{CC}}=0.36$~kJ/mol and $\sigma _{\mathrm{CC}}=0.34~$nm,
and combine these with oxygen-oxygen parameters $\varepsilon _{\mathrm{OO}%
}=0.65$~kJ/mol and $\sigma _{\mathrm{OO}}=0.32~$nm, using geometrical and
arithmetic means, respectively, to obtain their carbon-oxygen parameters.
They compute a potential of mean force by molecular simulation for a pair of
plates formed from these ``carbon'' particles. Figure. 1a of their paper
shows that the solvent contribution to the free energy for dissociated
plates is lower than that of associated plates. The plates are reasonably
large, so that the solvation energy is predominantly proportional to the
solute surface area exposed to the solvent. Choudhury and Pettitt's result for
the solvent contribution to the potential of mean force therefore shows that
the interactions used in their study produce a negative solvation surface free energy for the
hydrated plate.

\bibitem{Huang00b}
Huang, D.~M.; Chandler, D. {\em Proc. Natl. Acad. Sci. USA} {\bf 2000}, {\em
  97}, 8324.

\bibitem{Huang01}
Huang, D.~M.; Geissler, P.~L.; Chandler, D. {\em J. Phys. Chem. B} {\bf 2001},
{\em 105}, 6704.

\end{thebibliography}
\end{document}